# Towards Massive Machine Type Cellular Communications

Zaher Dawy, Walid Saad, Arunabha Ghosh, Jeffrey G. Andrews, and Elias Yaacoub[1]

## Abstract

Cellular networks have been engineered and optimized to carrying ever-increasing amounts of mobile data, but over the last few years, a new class of applications based on machine-centric communications has begun to emerge. Automated devices such as sensors, tracking devices, and meters – often referred to as machine-to-machine (M2M) or machine-type communications (MTC) – introduce an attractive revenue stream for mobile network operators, if a massive number of them can be efficiently supported. The novel technical challenges posed by MTC applications include increased overhead and control signaling as well as diverse application-specific constraints such as ultra-low complexity, extreme energy efficiency, critical timing, and continuous data intensive uploading. This paper explains the new requirements and challenges that large-scale MTC applications introduce, and provides a survey on key techniques for overcoming them. We focus on the potential of 4.5G and 5G networks to serve both the high data rate needs of conventional human-type communications (HTC) subscribers and the forecasted billions of new MTC devices. We also opine on attractive economic models that will enable this new class of cellular subscribers to grow to its full potential.

## 1 Introduction

The realization of smart cities in which homes, vehicles, and mundane objects are endowed with sensing and communication capabilities will accelerate towards 2020. MTC deployments and services are expected to grow exponentially and create a multi-billion dollar industry spanning a broad range of vertical sectors including transportation, utilities, health, environment, and security.

Machine-type deployments will generate many new and diverse forms of data traffic with varying requirements in terms of delay, per-link and total bit rate, reliability, energy consumption, and security/privacy. While traditional machine-type connectivity has relied on short-range wireless technologies such as Bluetooth, moving towards large-scale deployments requires broader interconnection capabilities which are best enabled by the wide-area coverage of cellular network infrastructures, especially with the current research, development, and standardization efforts towards a global 5G network. In this paper, we focus on "massive MTC", where a very large number of devices (at least 10x greater than the current cellular subscribers), with varying quality-of-service (QoS) requirements must connect to the cellular network. Realizing this in practice is contingent upon transforming the design, planning, and operation of cellular networks in terms of scalability and efficiency to handle the emerging challenges of the diverse and dense machine-type deployments.

---

[1] Z. Dawy (zd03@aub.edu.lb) is with the American University of Beirut (AUB), W. Saad (walids@vt.edu) is with Virginia Tech; Arunabha Ghosh (ghosh@labs.att.com) is with AT&T Lab; Jeffrey Andrews (jandrews@ece.utexas.edu) is with the University of Texas at Austin; Elias Yaacoub (eliasy@ieee.org) is with Strategic Decision Group (SDG) and Arab Open University (AOU).

The key motivation for deploying MTC over cellular is the forecasted massive number of MTC devices that will be deployed requiring significant aggregated data rates with global geographic area coverage. Cellular networks are the natural choice to handle a major part of this emerging traffic due to their already existing infrastructures, wide area coverage, and high-performance capabilities. Certainly, this will be also complemented with MTC connectivity over capillary networks and low power wireless local area networks. Therefore, MTC over cellular is not a matter of choice anymore; there are already MTC subscriptions supported by cellular operators worldwide, enhanced techniques are currently under standardization by 3GPP (e.g., see [1], [2]), industry forecasts predict steady exponential growth for MTC devices in the next five years, and MTC has been recommended as one of few central use cases for 5G development [3].

The open questions towards massive MTC deployment over cellular have to do with when it will happen, and how it should be designed and planned. To this end, there are major roadblocks that need to be addressed at both technical and economic levels. MTC services are a new paradigm for revenue generation for cellular operators, yet cannot be easily integrated into existing cellular subscription and activation models. One key challenge is to develop market models for understanding the interactions between mobile operators, service providers, vertical sector customers, and regulatory agencies. To this end, achieving appropriate equipment and service pricing will be key to the growth of MTC services. Also, the possibility of having mobile virtual network operators (MVNOs) that do not own the cellular spectrum, but rather lease it from spectrum owners, is another new twist to MTC over cellular.

The following are key questions that need to be addressed in order to enable large scale machine-type device deployments over cellular networks:

- How are MTC services different from standard HTC services? What are key use cases for MTC deployments over cellular?
- How will the operation/performance of cellular networks be impacted by the dense deployment of diverse MTC devices and services?
- Are the challenges to handle MTC over cellular real engineering challenges or will they be actually handled quite easily by already expected 4.5G and 5G cellular evolutions?
- If not, how to enhance/evolve/customize the design, planning, and operation of next generation cellular networks to meet the requirements of MTC devices/services? What potential techniques can be leveraged to intelligently manage or offload MTC traffic?
- How should cellular operators converge on attractive economic models and pricing schemes in collaboration with the various existing business constituents?
- How to provide incentives to MTC service providers to adopt cellular networks as a competitive connectivity option in terms of cost and quality versus other options such as proprietary or WiFi-based networks?

This paper focuses on these key questions, highlighting the main roadblocks and discussing possible ways forward. There is not a "one size fits all" solution and, thus, the techniques presented complement each other to handle the diverse classes and use cases of massive machine-type deployments over cellular infrastructure.

## 2   MTC Design Requirements and Their Impact

The requirements to support MTC services over cellular are inherently different and more diverse than traditional HTC services. MTC services add several challenging new constraints to the range of service requirements in cellular networks, as summarized in Table 1.

**Table 1: MTC versus HTC requirements in the context of cellular networks**

| Requirements | HTC over cellular | MTC over cellular |
|---|---|---|
| Uplink | Uplink is usually more lightly loaded and power-constrained | For many MTC applications, the main bottleneck; high signaling overhead and extreme power constraints |
| Downlink | The main bottleneck for high data rate services, since most traffic comes from the core network | Needs to be able to deep sleep, but wake up on command for network-initiated communication |
| Subscriber load | Relatively few (< 100) simultaneous devices per cell | Many (>> 100) simultaneous devices per cell with traffic uploading that can be event-triggered, periodic, or continuous |
| Device types | Relatively homogeneous, smart phones and data consumption devices like tablets | Extremely heterogeneous device landscape that includes environmental sensors, utility meters, wearable devices, and many unforeseen applications |
| Delay requirements | Defined service classes by 3GPP, vary between real-time conversational and best effort data | Very diverse delay requirements, ranging from emergency/time critical to very delay tolerant applications |
| Energy requirements | Flexible energy requirements due to the ability to recharge daily | Many ultra-low energy applications that require extreme power consumption measures |
| Signaling requirements | Signaling protocol overhead is not a concern and the design provides reliable mobility and connection management mechanisms | Application-dependent signaling protocols, with extremely efficient overhead signaling and contention resolution |
| Architectural requirements | Well-understood hierarchical cellular architecture with standardized interfaces between access and core network elements | Wide area coverage may require integration of data aggregators with multihop relaying; relaxed requirements for handover and roaming support |

Cellular operators will face performance quality challenges if they open their networks up to MTC service providers without taking necessary measures, at both the planning and operational levels, which comes with performance and cost implications. To demonstrate the potential impact of MTC devices on current cellular networks, we present sample results from an experimental case study conducted by the authors under realistic operational assumptions.

We used drive-testing equipment to quantify experimentally the signaling overhead of two standard MTC services over the air interface of a live 3G cellular network: (i) smart metering, which reflects static MTC traffic and (ii) low-speed vehicular sensing, which reflects mobile MTC traffic [4]. We utilized short message service (SMS) connections for messages of size less than 248 bytes (denoted as "SMS-type") and packet switched file uploads for larger message sizes (denoted as "email-type"). For each SMS-type and email-type connection established, we capture all the exchanged control signaling messages over the air interface in both uplink and downlink (connection management, measurement reports, security commands, mobility management, etc.), determine the size of each message, and then compute the signaling load. The signaling overhead is then calculated as the ratio of the signaling load to the sum of signaling and traffic loads.

The measured results are summarized in Figure 1 [4]. It is interesting to note that the signaling overhead is close to 100% for a range of application payload data sizes; this demonstrates the significant signaling overhead from MTC devices and motivates the need for efficient signaling reduction techniques. In general, the size of the downlink control messages is considerably larger than the uplink leading to higher signaling overhead. In addition, vehicular applications incur higher signaling overhead due to more frequent cell reselection and handover procedures. Finally, the signaling overhead corresponding to email-type connections notably exceeds the SMS-type thus rendering it inefficient for small MTC messages, e.g., for a 100 byte message, the uplink overhead is around 70% and 85% for SMS-type and email-type connections, respectively. Therefore, the application's payload raw data size and upload periodicity are prime factors that determine the best MTC connection type to minimize the signaling overhead.

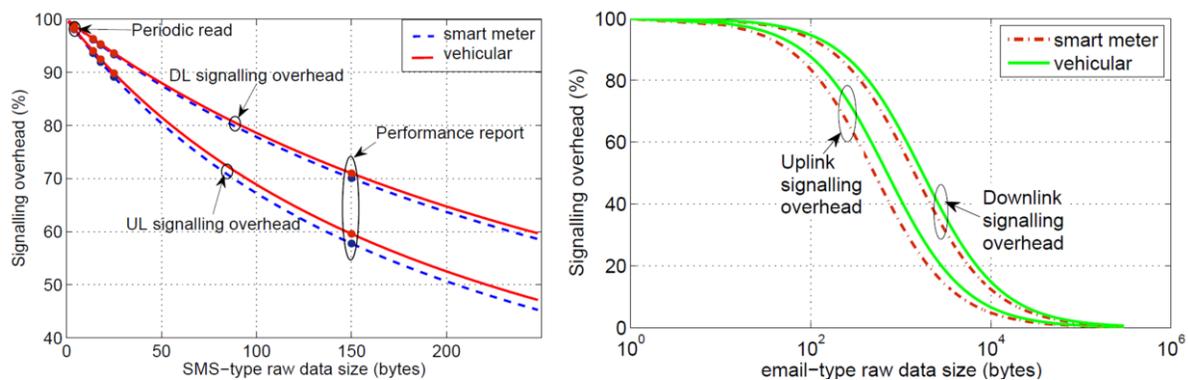

**Figure 1: Left: Signaling overhead for SMS-type MTC connections carrying short data for smart metering and vehicular sensing. Right: Signaling overhead for email-type MTC connections with larger data sizes.**

## 3   Key Technical Challenges and Design Strategies

Due to the diversity of MTC applications, it is not feasible to devise a single technical solution for all possible applications. As noted at the outset, we focus on massive MTC deployments. Therefore, MTC applications such as vehicle-to-vehicle communications, local indoor automation, or low density fixed sensor deployments are not a main focus since they can be efficiently handled via alternative technologies such as IEEE802.11p (for V2V), IEEE802.15.4, WiFi, and/or wired broadband. We now consider some of the envisioned MTC applications.

## 3.1 Overview of Use Cases and Requirements

Table 2 captures the main properties and requirements for various key use cases for M2M over cellular while highlighting as well effective strategies for optimized operation.

**Table 2: A summary of key use cases for MTC over cellular**

| Use cases | Deployment properties | Main requirements | Solution strategies |
|---|---|---|---|
| Smart utility metering | Fixed and known locations; medium to high density | Low rate; periodic traffic; small message sizes; uplink data | - Data aggregation<br>- Long sleep cycles<br>- Large link budget to cover deep indoor |
| Industrial automation | Fixed and known locations; low to medium density; hot spots | Low rate; periodic or continuous traffic; small message sizes; uplink data and *downlink control with critical timing* | - Data aggregation<br>- Multi-homed connections for reliability<br>- High priority bearer service for reliability |
| Video surveillance | Fixed and known locations; low to medium density | Medium to very high rate; *continuous traffic*; large message sizes; uplink data | - Data aggregation<br>- Event-based uploading<br>- Reservation based scheduling |
| Environmental sensing & actuation (forests, oceans, transportation, agriculture) | Fixed and arbitrary locations; high to massive density | Low rate; periodic or event triggered traffic; small message sizes; uplink data; downlink control; *ultra-low energy consumption*; low complexity | - Energy harvesting<br>- Deep sleep modes<br>- Low overhead uplink signaling protocols<br>- Timely paging for machine terminated traffic |
| Wearable sensing (health, activity, emotion, location) | Mobile and arbitrary locations; medium density | Low rate; periodic or event triggered traffic; small to large message sizes; *delay tolerant to time critical uplink data*; none to time critical downlink control; low energy consumption; low complexity | - Event-based uploading<br>- In-device caching with delay tolerant uploading<br>- WiFi offloading in indoor scenarios<br>- Multi-mode operation to support emergency situations |
| Vehicular sensing (vehicle indicators, safety, driving behavior, congestion levels, alarms, | Mobile and arbitrary locations; high density | Low to medium rate; periodic or event triggered traffic; small to very large message sizes; uplink data; | - Event-based uploading<br>- In-vehicle caching with opportunistic uploading and collection point offloading |
| surrounding) | | downlink control; *high mobility* | - Delay-based service priority for scheduling<br>- Multi-mode handover and cell reselection protocols |

Navigating through the various use cases and their properties and requirements, we can extract the following set of key challenges and opportunities:

- **Deployment-related challenges** including massive density, arbitrary locations, shadowed indoor and high mobility; opportunities include use cases with static fixed locations.
- **Traffic-related challenges** including continuous uploading, critical real-time uplink and/or downlink connections, and very large message size; opportunities include use cases with (very) delay tolerant traffic.
- **Energy-related challenges** including ultra-low energy consumption and high energy efficiency; opportunities include possibly automated battery charging.
- **Protocol-related challenges** including low overhead and scalable signaling with deep sleep modes, yet timely paging or network-initiated device triggering for time-critical downlink connections.
- **Complexity-related challenges** including transmitter/receiver design requirements, computational/storage capabilities, and device longevity through backwards compatibility, to maintain functionality over a long period of time even as cellular technologies evolve.

In the remainder of this section, we discuss effective strategies that can shape the design of next-generation cellular technologies and standards to support large-scale M2M deployments. These solutions complement ongoing activities in 3GPP standardization to develop new air interfaces to better support MTC devices, in addition to ongoing activities by IEEE 802.11ah Task Group to enhance wireless local area networks to support dense machine-type deployments [5].

### 3.2 Efficient Overhead Signaling Protocols and Procedures

As now established, a main technical challenge is the amount of overhead signaling generated by dense MTC devices. 3GPP has identified the causes for *uplink* signaling congestion due to a high number of devices trying almost simultaneously to attach to the network, to activate, modify, or deactivate a connection, or to perform cell reselection and handover procedures. There are several 3GPP proposals that present interesting ideas to reduce the signaling congestion problem [6], as well as basic research such as [7]. These include dedicated random access channel (RACH) formation, time-controlled policies with slotted access, prioritized connection rejection schemes, and network triggered polling. For example, in a smart metering MTC scenario, although the device is in a fixed location, it will report back measurement control messages to the network. In this case, it is possible to use spare bits in the connection request message to signal to the network that the device is stationary and, thus, disable all mobility-related signaling.

Another, and closely related, requirement for many MTC devices is extreme energy conservation. This includes many types of meters, sensors, and trackers, where recharging batteries may be difficult or impossible. This in turn requires most time to be spent in deep sleep modes, with either periodic or network triggered wake up. The advantage of periodic wake up is that it is simple; however, the period needs to be increased notably. For example, deep sleep times can now reach up to 10 min. in 3GPP Rel. 13, but this is still a few orders of magnitude too often for many MTC applications. Such infrequent wake ups conserve energy, but prohibit urgent or unpredictable control messages from the network; and such messages are essential for certain industrial automation, environmental monitoring, or mobile health MTC application scenarios. For example, if a gas leak is detected, all sensors need to be polled to figure out its location and possibly take immediate action. Thus, a more attractive option is to devise network-triggered mechanisms that can wake up sleeping devices quickly when needed, for example through passive reception of such triggers (similar to RFID) which is closely related to the emerging area of simultaneous wireless power and information transfer. Such wireless power transfers are implausible for high rate communication but could be used to provide MTC devices with energy to receive simple and infrequent wake-up signals while in deep sleep.

The notion of sleep mode should also be expanded to differentiate between radio communications sleep cycle and sensing/computing sleep cycle. That is, an MTC device can switch its communications interface off, but keep on sensing or an MTC device can switch both sensing and communications interfaces off depending on the use case requirements. Revisiting the gas leak example, the monitoring MTC devices can sense frequently with very low energy consumption and wake up only when a measurement exceeds a predefined threshold.

In addition to signaling overhead reduction, there are opportunities to exploit multiple radio access technologies (RAT) with technology-specific customizations for better support and improved performance efficiency. Ideas in this direction include recent 3GPP activities on a new cellular ultra-low complexity and throughput air interface [2], low cost LTE UE (user equipment) with reduced receive bandwidth to 1.4 MHz (Cat-0 UE category) [1], control and data plane separation over multiple cells that belong to the same or different RATs, and MTC device allocation to different RATs depending on service requirements and RAT capabilities.

### 3.3 Data Aggregation and Multihop Connectivity for MTC Devices

Hierarchical data aggregation is a key solution strategy to collect, process, and communicate data in MTC use cases with static devices, especially if the locations of the devices are known such as smart utility meters or video surveillance cameras. This requires the planning and deployment of aggregators that act as masters to collect information from pre-defined clusters of MTC devices over capillary short-range. The communication from the MTC devices to the aggregators can take place over multihop device-to-device (D2D) connections. D2D is expected to be an integral part of 4.5G and 5G technologies with direct benefits to MTC connectivity and use cases especially to overcome link budget problems for ultra-low power devices, e.g., see [3]. Relevant research challenges include MTC device clustering, transmission mode selection, in addition to interference and power management. Therefore, aggregators must be equipped with short range wireless interfaces to connect to MTC devices (such as LTE-Direct, WiFi-Direct, Bluetooth, or other proprietary solutions) in addition to cellular interfaces to relay traffic from MTC devices to the network. There is also potential for utilizing mmWave connectivity between MTC devices and aggregators for specific applications that require ultrahigh bit rates over very short distances; however, this requires addressing major research challenges related to the design of circuit components and antennas that meet the energy and complexity requirements of MTC devices.

For MTC use cases with static devices but *arbitrary* locations (e.g., deployment of sensors in a field), there is a need to devise self-organizing clustering mechanisms with either structured or arbitrary distribution of aggregator nodes. This is a complex problem especially with the support of multihop connectivity among the devices and the aggregators. An attractive solution approach is machine learning based spectral clustering where similarity matrices are determined based on MTC device information and utilized to perform clustering. Another more dynamic approach is coalitional game theory that allows not only clustering based on similarity, but also implementation of cooperative communication techniques that exploit the existing similarities.

Data aggregation has many potential benefits for many MTC over cellular use cases; (i) it reduces uplink signaling overhead by reducing the number of radio connections into any given receiver, (ii) it reduces transmit power by reducing the link distance, and (iii) it extends coverage range. To demonstrate the effectiveness of aggregation in reducing uplink signaling load, we present sample experimental results that quantify the performance tradeoffs of aggregation over a live 3G cellular network using a real-time drive testing setup. Data aggregation is modeled by the transmission of a large combined email-type message instead of multiple small messages from individual MTC devices. Figure 2 quantifies the signaling load as a function of the number of MTC devices connected to the aggregator [8], which we can see decreases as the aggregators become more numerous. Recent theoretical results derive bounds on the optimal number of hops, i.e. the number of levels of aggregators, finding that it is linearly proportional to the path loss exponent and outage target, and inversely proportional to the required SINR [9].

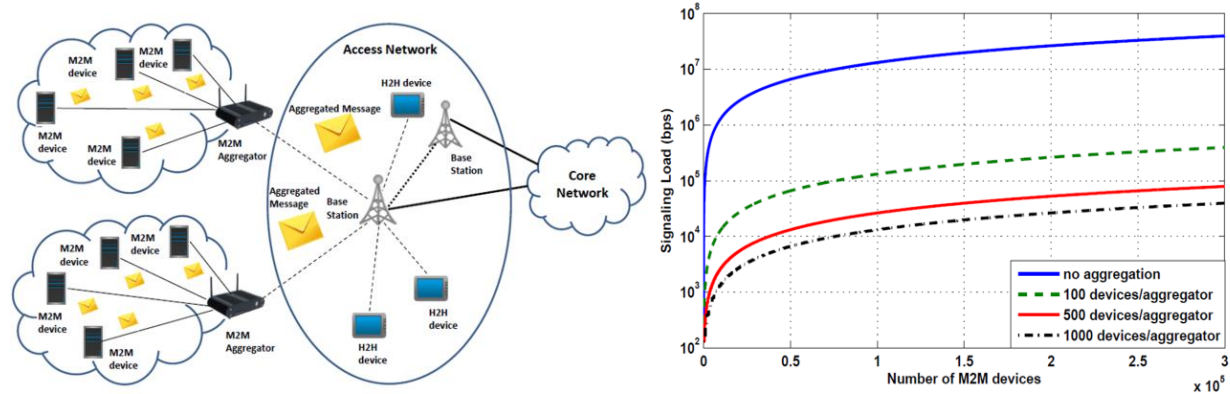

**Figure 2: Left: M2M aggregation scenario. Right: Uplink signaling load as a function of aggregation level and number of M2M devices for a given traffic load.**

The proliferation of MTC aggregators in high density geographic areas will lead to ultra-dense small cell like deployments with capacity, coverage, backhaul, and interference management requirements. In such scenarios, multi-RAT solutions with self-organizing features are expected to be highly effective, e.g., see [10] for a novel 3GPP-compliant architecture to reduce radio and core network congestion via home femtocells.

Finally, the role of an aggregator need not be limited to communications relaying but should be extended to support radio network intelligence (such as resource allocation, interference management, spectrum sharing, etc.) and local computing for higher efficiency. For example, aggregators can apply advanced data reduction techniques to compress the combined data before uploading, or apply feature extraction and event detection techniques to decide whether to upload the data or simply drop. This cross-layer fusion between communications and computing intelligence is pivotal for achieving an advanced degree of efficiency, robustness, and scalability.

### 3.4 In-Device Processing with Intelligent Resource Management

Distinguishing characteristics of some key MTC use cases are traffic predictability, event-based monitoring, and high delay tolerance. These facilitate the implementation of various dynamic in-device processing techniques in order to enhance the overall communications efficiency and reduce the load on the cellular network infrastructure.

- For smart utility meters or aggregators, the traffic is normally highly periodic over time and nearly deterministic in terms of data size. For such traffic, *reservation based scheduling* with lower signaling overhead (e.g., similar to semi-persistent scheduling for voice over LTE [11]) is attractive as it allows the cellular operator to dimension what radio resources are used and when from a large number of MTC devices per cell. For example, the MTC devices can be allocated the needed resources to upload their traffic in off-peak hours in coordination with the MTC application service provider.

- For MTC use cases such as vehicular sensing, sensors can collect data from the vehicle and its surrounding on a continuous basis with in-device caching and *long-term opportunistic scheduling* [12]. Depending on the delay tolerance of the collected data, the transmitting machine-type device can be allocated cellular resources depending on the current network load in the various cells that the vehicle is passing through. To control that the traffic caching delay does not exceed the service quality bound, the connection's service priority can be dynamically increased over time based on an adaptive algorithm. If the service delay bound is relatively flexible, then the cached data over a certain time window can be even offloaded over pre-deployed *data collection points* (e.g., WiFi access points) installed in specific locations that are either public (e.g., street crossings) or private (e.g., parking lots).

- For use cases such as environmental sensing via static MTC devices or wearable sensing via mobile MTC devices, a main application is monitoring with *event-triggered communications* (e.g., emergency health state or alarming temperature level). In this case, the device will need to mainly upload data, which could be highly urgent, as soon as an event is detected. Thus, the cellular operator needs to provide a new high priority uplink service for critical MTC uploading; this can directly build on the existing 3GPP emergency call procedures. The need for critical timing in MTC communications extends to downlink cellular, e.g., for sensor control in emergency situations; this can be achieved via network-triggered device wake-up and control mechanisms as discussed in Section 3.2.

- In scenarios with limited coordination between the cellular operators, MTC service providers, and/or MTC customers, there is a need to develop *adaptive machine learning* algorithms that can accurately predict the nature of the traffic and transmission patterns. This information will then be used to develop *self-organizing resource management mechanisms* that can exploit the learned data to optimize performance without disrupting HTC connections quality. A promising approach in this direction is to exploit low-complexity and distributed mathematical frameworks, such as matching theory [13], which enable optimized allocation of resources to devices, while accounting for locally learned data.

## 3.5 Meeting Energy and Longevity Requirements

In this section, we focus on two of the most demanding operational constraints for MTC that cannot be adequately addressed via incremental enhancements: ultra-low energy consumption and device longevity. These challenges are naturally coupled with complexity constraints at the device level in terms of communications, computing, and storage capabilities.

Due to battery capacity limitations, *energy harvesting* techniques can provide MTC devices with the needed energy to maintain connectivity and functionality over long periods of time without the need for recharging. The choice of energy harvesting technique (e.g., solar, vibration, heat) will depend on the MTC device's hardware design, function, operational, and deployment conditions. For example, health-related wearables could harness the body's heat or vibrations, whereas an outdoor sensor could use a very small solar panel. Although energy harvesting is currently a hot topic for wireless applications in general, it seems that massive MTC is a particularly attractive application of energy harvesting since the amount of total required energy is fairly small, and the need to avoid recharging is so great.

*Longevity* is an important and usually overlooked operational challenge that is quite specific to MTC over cellular. Unlike smartphones which are replaced every 2-3 years on average and are rarely still in use after 5 years, utility meters and other MTC sensors routinely require a guaranteed lifetime of fifteen years or more before replacement. The business model becomes unattractive if an MTC service provider has to replace devices as new cellular technologies are standardized for the faster churning commercial HTC applications. We suggest two main approaches to guarantee longevity. The first is to have cellular operators provide guarantees to maintain backward compatibility for a given period of time, e.g., 20 years. Although this solution is transparent to the MTC devices, it greatly restraints future innovation and flexibility, and could have damaging long-term effects. However, these effects could possibly be reduced by introducing new MTC technologies on different bands, while maintaining the backwards compatibility on legacy bands.

The other approach is to develop MTC devices with software defined radio (SDR) capabilities to facilitate a certain degree of re-configurability via software updates. However, this is still complex since the devices will be designed without knowing how technology standard releases will evolve 10 to 20 years later; in addition, the devices might have complexity constraints to handle re-configurable hardware. Therefore, the objective should be to converge at some middle ground that allows for a certain level of reconfiguration combined with a certain level of backward compatibility from the operators, e.g., physical layer functions can be fixed whereas MAC and network layer protocols and procedures can be software configurable.

## 4 Economics of MTC Deployments

The deployment of MTC services over cellular networks introduces new economic challenges that can significantly differ from existing models. Indeed, the diversity of the MTC service offering will require rethinking to the way existing cellular network economic models operate as discussed in this section.

### 4.1 Overview of Economic Considerations and Challenges

A summary of the key economic challenges for MTC deployment is provided in Table 3. Hereinafter, an MTC customer is defined as a user who subscribes to MTC services provided directly by an operator while an MTC provider is a dedicated third-party that offers MTC services which is eventually delivered over the operator's network.

Table 3: MTC versus HTC economic aspects in the context of cellular networks

| Economic aspects | HTC over cellular | MTC over cellular |
|---|---|---|
| Market players | Operators; MVNOs; regulatory bodies | Operators; MTC providers/vertical sector customers; MVNOs; regulatory bodies |
| Market model | Hierarchical model with pre-defined leader-follower structure | Hierarchical model with varying leader-follower structure |
| Role of operators | Deliver connectivity and wireless services | Deliver own and third-party MTC services on top of connectivity and existing services |
| Pricing plans | Well-defined pricing plans | MTC service dependent; smart data pricing; net neutrality questions |
| Data caps | "Standardized" data caps per individual subscriber | Caps not viable for all MTC services; need for aggregate data caps per service |
| Profit or cost sharing | Not common | Cooperation between operators and MTC providers |

Cellular operators and mobile virtual network operators (MVNOs) will be required to deliver MTC services to their own customers as well as customers of third-party MTC providers. MVNOs are virtual operators that do not own spectrum or infrastructure but rather lease it from the owner. Operators and MVNOs must consider how to build economically profitable partnerships with MTC providers such as a transportation service provider or a smart grid utility while building their own market base. The relationship between operators and MVNOs must also be revisited in light of MTC deployments. For example, operators might also consider MVNOs as a cost-effective way to offload MTC traffic depending on how machine-type services will grow and how such growth will impact the technical operation of the network. A key economic question here would be how to manage the lease of infrastructure and spectrum to MVNOs for MTC purposes.

In addition, the role of regulatory bodies in ensuring a gradual and effective deployment of MTC services must be analyzed. Examples of such involvement can include questions on whether such bodies might want to provide certain incentives, monetary or in terms of spectrum, for operators to expedite the deployment of MTC services. How such incentives must be designed will constitute an important challenge for MTC deployment.

Last but not least, there will be a need for cooperation and possibly revenue sharing mechanisms between operators and third-party MTC providers. Cooperation between operators themselves may also be needed to regulate and better manage the expected massive deployments. Such cooperation and profit sharing mechanisms are minimal and even non-existent in classical cellular markets. The role of MTC customers in driving such profit sharing mechanisms can also be critical and must be analyzed.

## 4.2 Basic Study of an MTC Market Structure

The MTC market structure will follow closely the hierarchical model of existing HTC markets. In such markets, a leader-follower structure exists in which the economic decisions made at an upper hierarchy (e.g., operators) can propagate to the lower hierarchies (e.g., customers) and vice versa.

In MTC, in contrast to the static HTC market structure, the exact leader/follower role can be dynamic depending on the scenarios. For example, when leasing infrastructure, operators are naturally leading the decisions of MVNOs, MTC providers, and MTC customers. However, when new services are launched by an MTC provider, such decisions will lead and impose changes to the operators' own network. This dynamic hierarchical model motivates a game-theoretic approach to analyze the MTC market. A natural tool is the framework of Stackelberg game, which must handle two unique challenges: 1) a multi-level, multi-leader, multi-follower structure as opposed to the classical two-level, single-leader structure and 2) dependence of the objectives of the players on externalities such as the technical decisions made. Indeed, the interplay between the network economics and the technical challenges must be properly factored in for effective MTC deployment.

To provide a basic insight on such a framework, we investigate, based on our work in [14], a multi-leader, multi-follower structure for studying the economics of deploying new services in a system having a large population of customers with heterogeneous requirements. Here, competing service providers (SPs) aim to deploy similar MTC services. The goal is to analyze optimal prices and customer selections. Using this framework, as shown in Figure 3, we can analyze a duopoly market in which SP 1 offers two services: service 1 and service 2 and another SP 2 offers only service 1. SP 1 is evaluating how much to economically invest in service 2.

Each service is associated with a unique spectrum, hence the objective functions depend exclusively on each service's own load. We use $G_0$ to represent the maximum QoS offered by SP 1's service 2; $G_0$ depends on the economical effort needed to deploy service 2 by SP 1 (to deploy a higher QoS service, one needs a higher $G_0$). Then, we let $G_0$ vary from 0 (i.e., no new service) to 5 (two services by SP) thus evaluating the SPs' revenues. From Figure 3, we can see that the income of SP 1 is boosted by the new service only when $G_0 > 3$. When $G_0 < 3$, SP 2 can approximately limit the loss of users by decreasing the price asked for the service. In this case, the effect of the implementation of the new service by SP 1 is to slightly decrease the profits of SP 2 (therefore improving the utilities of the users), instead of increasing the revenues of SP 1. Clearly, within the MTC market structure, pricing will play an important role as discussed next.

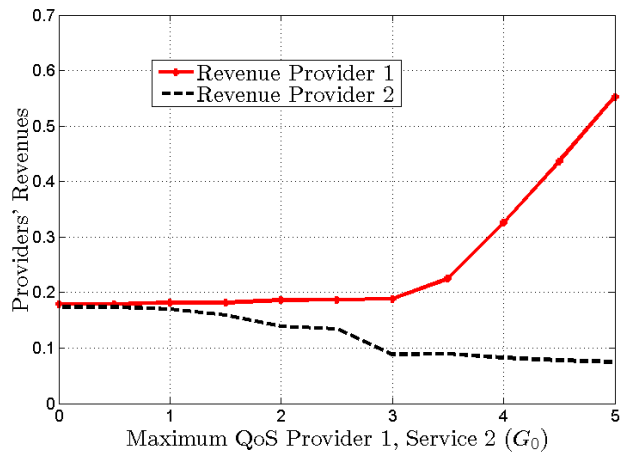

**Figure 3: On M2M over cellular economics: Sample results for a case study with two service providers.**

### 4.3 Pricing Plans and Data Caps for MTC Services

In a classical cellular network, well-defined pricing plans and bundles already exist. However, incremental changes to such plans will not be sufficient to handle the MTC market. For example, if an operator delivers services for both third-party MTC providers and its own MTC customers, it must consider how to price these two different offerings so as to maximize revenues, ensure service quality, and avoid competing with itself by inadvertently pushing customers to prefer the third-party providers. This also brings in interesting questions of net neutrality since operators might not be able to provide preferential services for their own customers at the behest of customers of the third-party providers.

The pricing plan will also depend on the type of the MTC service itself. For instance, critical MTC services, which require higher QoS and maintenance, may need to pay a premium as opposed to massive, non-critical services (such as smart utility meters). In addition, owing to the heterogeneity and diversity of machine-type devices and services, the usage of smart data pricing will become more prominent [15]. Such smart data pricing policies can include time-dependent and location-dependent pricing which can be tailored to specific business scenarios. For example, non-critical MTC services can benefit from time-dependent pricing plans that incentivize them to transmit their data during non-peak hours thus also contributing to reducing the congestion of the system, as discussed in Section 3.3.

One additional challenge is the notion of data caps and volume bundles. For regular subscribers, per-user data caps are common and widely used. In contrast, for MTC services, defining per-device data caps may not be viable, particularly for critical services or for services that rely on low-data rate, small devices, such as sensors. In such cases, one may consider introducing aggregate data cap models. Such models can be set on a per MTC provider case, thus providing a fixed price based on the total amount of data that a certain provider can use within a given time period or even per location. Determining these caps will also be driven by the introduction of accurate models for MTC traffic.

Beyond pricing services, other economic challenges include investigating how to price and offer various machine-type devices, how to provision and allocate blocks of SIM cards to MTC customers with flexible activation/deactivation options, and how to perform automated on-demand device configuration.

Clearly, the massive deployment of MTC over cellular networks poses important and non-trivial economic considerations that must be jointly considered along with the technical challenges for paving the way towards the successful realization of a key requirement as part of the 5G vision.

# 5  Conclusion

Cellular technologies have inherent advantages to play a central role in serving the exponentially growing number of machine-type devices, due to their wide area coverage and high capacity characteristics. Yet, this is contingent on addressing major challenges at both technical and economical levels. In this article, we present a set of effective design strategies tailored to various MTC use cases that can enhance the design of next generation cellular technologies. In addition, we cover key economic considerations for facilitating the deployment of MTC over cellular from a business perspective, and discuss a basic study of an MTC market structure.

## Acknowledgements

This work was made possible by NPRP grant 4-353-2-130 from the Qatar National Research Fund (a member of The Qatar Foundation). The statements made herein are solely the responsibility of the authors.